%% file: Compact-objects-revised-v1.tex
\documentclass[a4paper,11pt]{article}
\usepackage{aaskaiid}
\usepackage{orcidlink}
\usepackage{threeparttable}
\usepackage{subcaption}
\usepackage[encapsulated]{CJK}

\title{Compact Objects Revealed by SKA and SKA-VLBI}
\ShortTitle{Compact Objects}

\author[1]{Lin, Z. H. \orcidlink{0009-0006-0392-6345}}
\author[1]{Li, Y. J. \orcidlink{0000-0001-7526-0120}}
\ShortName{Lin et al.} 
\author[1]{Hao, C. J.\orcidlink{0000-0002-6820-198X}}
\author[1]{Li, J. J.\orcidlink{0000-0002-3338-8465}}
\author[1]{Dong, Y. W.\orcidlink{0009-0000-5512-9109}}
\author[2, 3]{Liu, D. J.\orcidlink{0009-0001-9837-9455}}
\author[1]{$^{\ast}$Xu, Y. \orcidlink{0000-0001-5602-3306}}

\affiliation[1]{Purple Mountain Observatory, Chinese Academy of Sciences, Nanjing 210023, China}
\emailAdd{xuye@pmo.ac.cn, linzh@pmo.ac.cn, liyj@pmo.ac.cn, cjhao@pmo.ac.cn, }
\affiliation[2]{College of Science, China Three Gorges University, Yichang 443002, China}
\affiliation[3]{Center for Astronomy and Space Sciences, China Three Gorges University, Yichang 443002, China}

\abstract{
Compact objects represent a crucial interdisciplinary frontier between astronomy and fundamental physics. The SKA/SKA-VLBI, with exceptional sensitivity (at $\mu$Jy levels) and ultrahigh positional precision (at $\mu$as levels), will enable direct, precise measurements of orbital dynamics in compact objects including black holes and neutron stars. This facility is expected to achieve at least three breakthroughs: (1) Developing effective methodologies for detecting and identifying black hole–neutron star binaries to construct observational catalogues, thus 
advancing investigations into the equation of state of ultra-dense nuclear matter and strong-field relativistic effects; (2) Determining critical parameters such as orbital elements and component masses in compact binaries, yielding insights into stellar structures and evolutionary mechanisms under extreme conditions; (3) Identifying intermediate-mass black holes and measuring their masses to deepen understanding of black hole formation and evolution.}

\begin{document}
\include{journal-names}
\maketitle
\begin{CJK*}{UTF8}{gbsn}
	
\section{Introduction}\label{sec:introduction-compact-object-astrometry}

Compact objects, including white dwarfs (WDs), neutron stars (NSs), and black holes (BHs), serve as critical laboratories for nuclear and particle physics, establishing a pivotal interdisciplinary frontier between astronomy and fundamental physics. Compact binaries (CBs) comprising these objects include NS X-ray binaries (NSXBs, a subclass of X-ray binaries, hereafter XBs), BH XBs (BHXBs), BH-NS binaries (BH-NSs), and systems hosting pulsars (PSRs) or WDs. Such systems are essential for both compact object and binary star research, advancing investigations into stellar structure and evolution under extreme conditions, environmental effects originating from star clusters or galactic environments, verification of relativistic phenomena, and temporal variations of the gravitational constant \citep{Ding+2023}. In particular, BH-NSs deliver multifaceted insights: constraining massive binary evolution scenarios, probing ultra-dense nuclear matter equation of state, elucidating gravitational-wave source formation mechanisms, and testing strong-field relativity effects \citep{Shao-Li2018}. These capabilities establish BH-NSs as essential astrophysical benchmarks. 

The existence of stellar-mass BHs (sBHs; $\lesssim$100 $M_{\odot}$) and supermassive BHs ($\gtrsim$10$^{5}$ $M_{\odot}$) is well established. However, direct observational confirmation of intermediate-mass BHs (IMBHs; $\gtrsim$100--10$^{5}$ $M_{\odot}$) remains scarce. Identifying and characterizing IMBHs will complete the BH mass spectrum, substantially advance fundamental comprehension of BH formation and evolution, and catalyze gravitational-wave astronomy development \citep{Greene+2020}.

Precise mass measurement of compact objects represents a principal research objective and methodological challenge in the field. Advancements in multi-messenger astronomy have diversified compact object mass-determination approaches, most prominently gravitational-wave detection \citep[e.g.,][]{Abbott+2023}, electromagnetic spectroscopy and/or photometry \citep[e.g.,][]{Jayasinghe+2021}, PSR timing \citep[e.g.,][]{Kramer+2006}, and astrometric techniques.

Astrometric techniques permit direct monitoring of orbital motions in binaries to infer component masses. Radio observing is the only known technique capable of directly tracking the orbits of compact objects in CBs; remarkably, just two instances have been resolved to date: (1) Very Long Baseline Array monitoring of sBH positional variations in the BHXB Cyg X-1 --- projected onto the photosphere where jet emission escapes --- enabled \citet{Miller-Jones+2021} to resolve the sBH's orbital motion and determine its mass as $21.2 \pm 2.2$ $M_{\odot}$; (2) \citet{Miller-Jones+2018} derived the orbital elements for the PSR binary, PSR B1259-63, through combined PSR timing and orbital monitoring of the PSR companion with the Australian Long Baseline Array.

Current sensitivity limitations, yielding signal-to-noise ratios of $\sim$10--20 and astrometric precisions of $\sim$40--100 $\mu$as, represent the primary challenge in monitoring compact objects. The SKA-VLBI will enhance sensitivity of current VLBI by approximately an order of magnitude (reaching the $\mu$Jy level) and advance astrometric precision to the $\mu$as level (potentially achieving 1 $\mu$as), thus establishing unparalleled capabilities to 
systematically monitor orbital motions in CBs and radio stars orbiting massive objects \citep{Li+2024-SKA-VLBI}, such as IMBHs. This work conducts a comprehensive examination of SKA-Mid and SKA-VLBI capabilities in: (1) detecting CBs and monitoring their orbital motions; (2) identifying IMBHs and measuring the orbital dynamics of their surrounding objects; evaluating potential paradigm-shifting progress in CB and IMBH research.

\section{Specifications of SKA and SKA-VLBI}\label{sec:specification-compact-object-astrometry}

Table \ref{tab:SKA-specification-compact-object-astrometry} presents the specifications of the SKA-Mid and SKA-VLBI \citep{Paragi+2015, Li+2024-SKA-VLBI}. The dynamic range, DR, is estimated as the flux density of the target source divided by sensitivity, $\Delta S$. The maximum DR, $\mathrm{DR_{max}}$, is set at 100:1 throughout this work. 

\begin{table*}[htbp]
\vspace{-0.2cm}
	\centering
		\fontsize{8}{7.9}\selectfont
		\setlength\tabcolsep{3pt}
	\renewcommand{\arraystretch}{1.1}
	\begin{threeparttable}
		\caption{Sensitivity and Astrometric Precision \label{tab:SKA-specification-compact-object-astrometry}}
		\begin{tabular}{lcccccc|lcccccc}
			\hline \hline
			Stages & $\nu$ & $b$ & $\Delta \nu$ & $\Delta S$  & $\theta$ & $\Delta \theta$ & Stages & $\nu$ & $b$ & $\Delta \nu$ & $\Delta S$  & $\theta$ & $\Delta \theta$ \\
			& (GHz) &  (km) & (GHz) & ($\mu$Jy beam$^{-1}$) & (mas) & (mas) & & (GHz) &  (km) & (GHz) & ($\mu$Jy beam$^{-1}$) & (mas) & (mas) \\
			\hline
			AA2      & 1.6  & 36  & 0.32 & 8.7 & 1100 & 5.5 & AA4      & 1.6  & 160 & 0.32 & 4.1 & 240 & 1.2\\
		         	& 8.0   & 36  & 0.80 & 6.9 & 220 & 1.1 &          & 8.0  & 160 & 0.80 & 3.3 & 48 & 0.24	 \\
		        	& 15.0  & 36  & 0.80 & 11.1 & 110 & 0.55 &          & 15.0 & 160 & 0.80 & 5.0 & 26 & 0.13 \\
			\hline
			AA$\ast$ & 1.6  & 36  & 0.32 & 7.0 & 1100 & 5.5 & SKA-VLBI & 1.6  & 10,000 & 0.128 & 4.9 & 3.9 & 0.02 \\
			         & 8.0  & 36  & 0.80 & 5.6 & 220 & 1.1 &           & 8.0  & 10,000 & 0.256 & 7.3 & 0.77 & 0.004 \\
		             & 15.0 & 36  & 0.80 & 8.4 & 110 & 0.55 & & 15.0 & 10,000 & 0.256 & 10.8 & 0.41 & 0.002  \\                                                      
			\hline
		\end{tabular}
		\begin{tablenotes}
		\vspace{2ex}
			\item \parbox{\linewidth}{
                  \setlength{\parindent}{-2em}
                  \setlength{\leftskip}{-1.5em}
                  \setlength{\rightskip}{0pt plus 0pt minus -2em} \noindent\hspace*{-0.4em} \noindent Note. 
                  The columns present: frequncy, $\nu$; baseline, $b$; bandwidth, $\Delta \nu$; sensitivity, $\Delta S$, for integration time of 10 min; full width at half maximum beam size, $\theta$; and position uncertainty (astrometric precision), $\Delta \theta = \theta/(2\cdot \mathrm{DR_{max}})$. These parameters are given for SKA-Mid AA2 and AA$^{\ast}$ (excluding dish SKA008), AA4 and SKA-VLBI.}
		\end{tablenotes}
	\end{threeparttable}
	\vspace{-0.4 cm}
\end{table*}

\section{Compact Binaries}\label{sec:flux-density-monitoring-compact-object-astrometry}

\subsection{Flux Density of Compact Binaries}\label{sec:flux-density-monitoring-1-compact-object-astrometry}

CBs exhibiting radio emission primarily include XBs (including BHXBs and NSXBs), PSR binaries (including double PSRs), BH-NSs, and novae. NSXBs evolve into PSR binaries, with some systems ultimately forming double PSRs \citep{Marchant-Bodensteiner2024}; intermediate evolutionary stages may include accreting millisecond X-ray PSRs (AMXPs) and transitional millisecond PSRs (tMSPs). Detected BHXBs, NSXBs, AMXPs, and accretion-powered tMSPs exhibit flat radio continuum spectra \citep[$S_{\nu} \propto \nu^{\alpha}$ with $\alpha \sim 0.0$--0.7;][]{Tetarenko+2016}, while PSRs and rotation-powered tMSPs display steep spectra. Higher frequencies provide improved astrometric precision and therefore take precedence. Therefore, the observing frequency is set at $\sim$1.6 GHz for PSR-containing binaries (for broad astrometric measurements) and at $\sim$15 GHz for other CB types.

\begin{figure*}[ht]
    \centering
    \vspace{-0.2cm}
    \subfloat[]{\hspace{-0.8 cm}\includegraphics[height=0.29\textwidth]{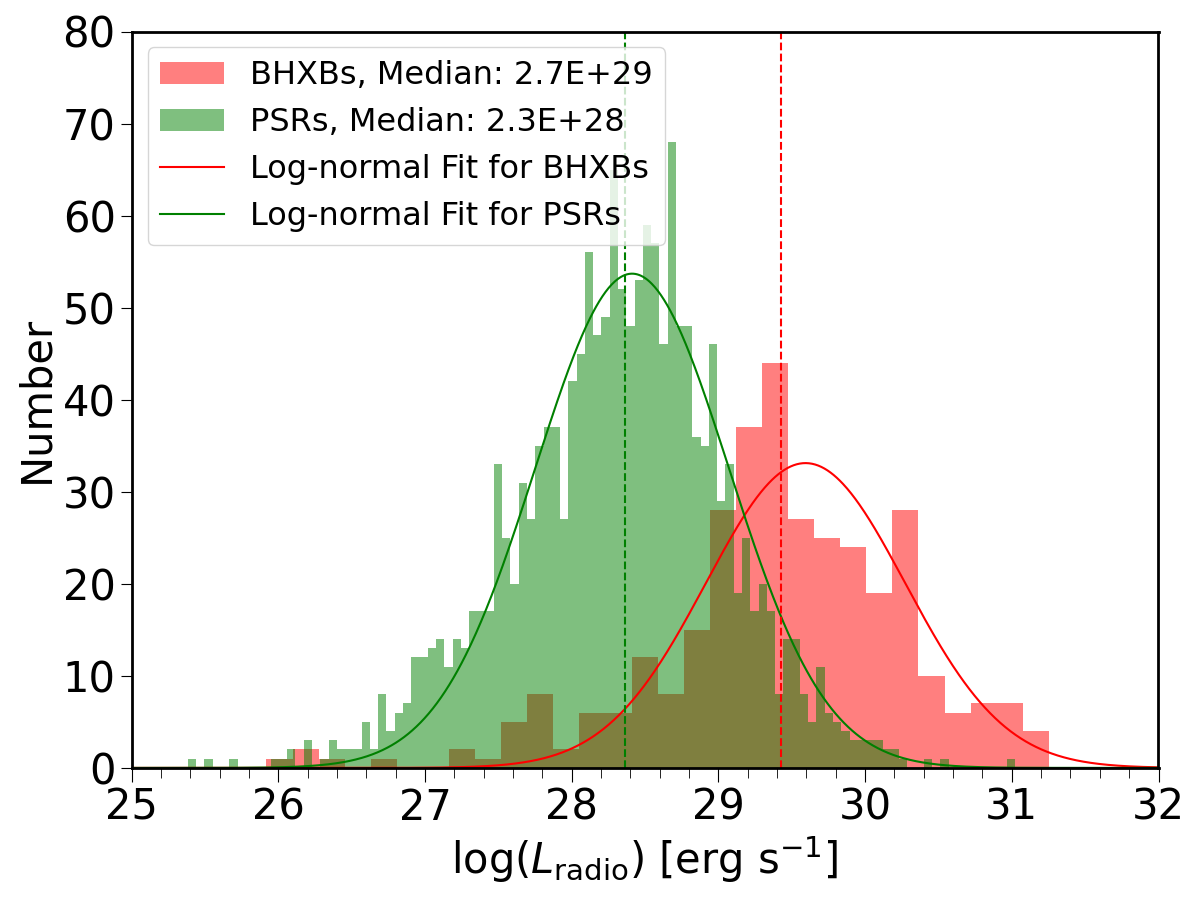}\hspace{1.6 cm}}
    \subfloat[]{\includegraphics[height=0.29\textwidth]{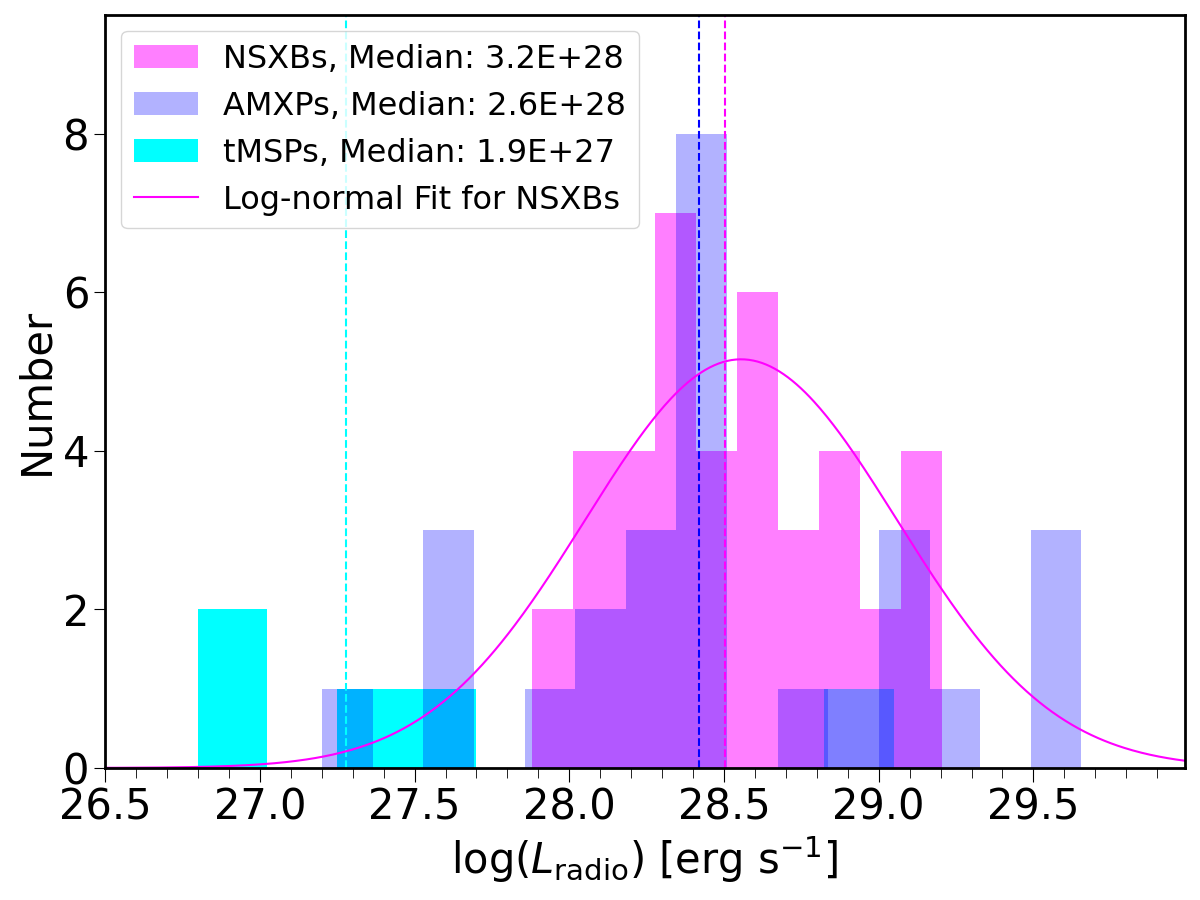}}
    \vspace{-0.4 cm}
	\caption{Radio luminosity, $L_{\mathrm{radio}}$, distributions of: (a) PSRs and BHXBs; (b) NSXBs, AMXPs, and accretion-powered tMSPs. The reference frequency is 1.4 GHz for PSRs and 5 GHz for others. Vertical dashed lines indicate the median $L_{\mathrm{radio}}$ values. Solid colour lines represent the log-normal fits.}
    \label{fig:flux-density-compact-object-astrometry}
\end{figure*}

Based on the ATNF PSR Catalogue \citep{Manchester+2005}, PSRs exhibit a median luminosity of $\sim${}$2.3 \times 10^{28}$ erg s$^{-1}$ @ 1.4 GHz (Figure \ref{fig:flux-density-compact-object-astrometry}). This corresponds to a median flux density of $\sim$13.7 mJy/$d^2$, where $d$ is the distance in kpc. In this work, we assume a similar flux density at $\sim$1.6 GHz. Despite their typically steep spectra, a few PSRs are still detectable at higher frequencies \citep[e.g., parallax determination of PSR J0437-4715 at 8.4 GHz;][]{Deller+2008}. High-frequency observations are preferred for astrometric measurements whenever possible.

\citet{Bahramian-Rushton2022} compiled radio and X-ray luminosity for BHXBs, NSXBs, AMXPs, and accretion-powered tMSPs. Figure \ref{fig:flux-density-compact-object-astrometry} presents the distribution of their radio luminosity at 5 GHz. The median flux densities of them are $\sim$45, $\sim$5.3, $\sim$4.3, and $\sim$0.3 mJy/$d^2$, respectively. Given their flat radio spectra, we set the flux densities at 1.6, 8, and 15 GHz equal to the value at 5 GHz.

\citet{Gulati+2023} conducted a systematic search for novae and estimated their 1 GHz flux densities at varying ejected masses. With an ejecta mass of $5 \times 10^{-6}$ $M_{\odot}$, the flux density reaches $\sim$0.6 mJy/$d^2$. The $\alpha$ of novae can approach $\sim$1, yielding a flux density of $\sim$9 mJy/$d^2$ at 15 GHz.

\subsection{Compact Binaries Survey}\label{sec:survey-monitoring-compact-object-astrometry}

The population of low-mass BHXBs in the Milky Way was estimated to be $\sim$10$^4$--$10^8$ \citep{Tetarenko+2016}.
With the NSXB-to-BHXB ratio estimated at $\sim$1--9 \citep{Jonker+2011}, significant populations of both systems remain undetected in the Milky Way. The initial CB candidate samples will comprise:  (1) X-ray point sources or candidates at other bands; (2) employing the forthcoming Gaia DR4 and future releases to search for CB candidates by detecting the orbital motions of non-compact companions.

The primary instrument for CB searches will be the SKA. With an integration time, $\tau$, of 10 sec (enabling observations of $\sim$10,000 sources across three bands in 100 hr), AA2 achieves a sensitivity of $\sim$53--86 $\mu$Jy at 1.6, 8.0, and 15 GHz. Using the median flux density and assuming a signal-to-noise ratio of 10, AA2 can detect BHXBs, NSXBs, AMXPs, and accretion-powered tMSPs out to distances, $d_{\mathrm{f}}$ ($\propto \tau^{1/4}$), of $\sim$7.2, $\sim$2.5, $\sim$2.2, and $\sim$0.6 kpc, respectively. Thus, starting from AA2, the SKA can conduct surveys for CBs by measuring spectral flatness.

Multi-epoch observations of a CB candidate will enable measurements of variability and proper motions. Variability constitutes a fundamental radiative characteristic of CBs. Beyond the spectral index and variability, further emission properties such as polarisation and radio jets merit scientific investigation. Proper motion measurements can test whether CB candidates are located within the Milky Way, thereby refining the sample \citep[e.g.,][]{Atri+2022}.

These CB surveys will yield critical insights into compact object accretion physics, high-energy radiation mechanisms, and binary evolution. Notably, combined with then-extant PSR and nova samples, they will supply an extensive candidate pool for orbital monitoring using SKA/SKA-VLBI.

\subsection{Monitoring Orbital Motion of Compact Binaries}\label{sec:monitor-monitoring-compect-object-astrometry}

Astrometry via parallax determination and orbital monitoring is one of the principal methods for measuring the physical parameters of CBs. While there are dozens of existing parallax determinations \citep{Li+2024-SKA-VLBI}, orbital monitoring, which requires higher sensitivity and precision, has only two valid instances \citep{Miller-Jones+2018, Miller-Jones+2021}, a domain in which SKA-VLBI excels.

A full astrometric solution includes the reference position and proper motion in both coordinates (e.g., $\alpha_0$, $\delta_0$ in the Equatorial Coordinate System; $\mu_\alpha\cos\delta$, $\mu_\delta$), the parallax, $\varpi$, and the seven orbital elements (orbital period, $T$, epoch of periastron passage, $T_0$, eccentricity, $e$, inclination, $i$,  semi-major axis of the primary, $a_1$, argument of periastron, $\omega$, and longitude of the ascending node, $\Omega$). Multi-epoch monitoring of compact objects in CBs is essential for deriving reliable astrometric solutions.

The astrometric mass, $f_{\rm M}$, can be determined from
\begin{equation}\label{equ:useless_0-compact-object-astrometry}
	f_{\rm M} = \left(\frac{a_1}{\varpi}\right)^3 \left(\frac{1}{T}\right)^2 = M_2 \left(\frac{M_2}{M_1 + M_2}\right)^2, 
\end{equation}
where $M_1$ and $M_2$ are the masses of the primary and companion in $M_{\odot}$, respectively. $a_1$ and $\varpi$ are both in units of mas, and $T$ in units of yr. A fixed integration time of 10 min is assumed throughout this subsection. The corresponding sensitivity, $\Delta S$, in $\mu$Jy; astrometric precision, $\Delta \theta$, in mas at $\mathrm{DR_{max}}$; and beam size, $\theta$, in mas are adopted from Table \ref{tab:SKA-specification-compact-object-astrometry}. Given that the parameter $a_{1}$ can only be measured if it exceeds the positional uncertainty, along with the median flux density of the target source, $S$, in $\mu$Jy (Section~\ref{sec:flux-density-monitoring-1-compact-object-astrometry}), the maximum detectable distance, $d_{\mathrm{max}}$, in kpc can be estimated from
\begin{equation}\label{equ:dmax-compact-object-astrometry}
	d_{\mathrm{max}} = \min\left(\left(\frac{8 f_{\rm M} S^3 T^2}{\theta^3 \Delta S^3}\right)^{-1/9}, \; \left(\frac{ f_{\rm M} T ^2}{\Delta \theta^3}\right)^{-1/3}, \; \left(\frac{ 2 S}{5\theta \Delta S}\right)^{-1/3}, \; \frac{1}{5\Delta \theta}\right), 
\end{equation}
where the last two terms represent the parallax limit.

For a 15 $M_{\odot}$ sBH in a BHXB with $M_{2} = 1$ $M_{\odot}$ and $T = 30$ days, Equation (\ref{equ:dmax-compact-object-astrometry}) yields $d_{\mathrm{max}}$ of $\sim$0.05, $\sim$0.23, and $\sim$8.46 kpc for AA$^{\ast}$, AA4, and SKA-VLBI, respectively. 
For a 1.4 $M_{\odot}$ PSR in a BH-NS, a PSR binary or a rotation-powered tMSP with $M_{2} = 8$ $M_{\odot}$ and $T = 10$ days, the $d_{\mathrm{max}}$ values for AA$^{\ast}$, AA4, and SKA-VLBI are $\sim$0.03, $\sim$0.14, and $\sim$6.15 kpc, respectively. Results indicate that AA$^{\ast}$ and AA4 can monitor nearby CBs, whereas SKA-VLBI significantly enhances $d_{\mathrm{max}}$ values.

\begin{figure*}[ht]
    \centering
    \vspace{-0.2cm}
    \subfloat[]{\includegraphics[height=0.3\textwidth]{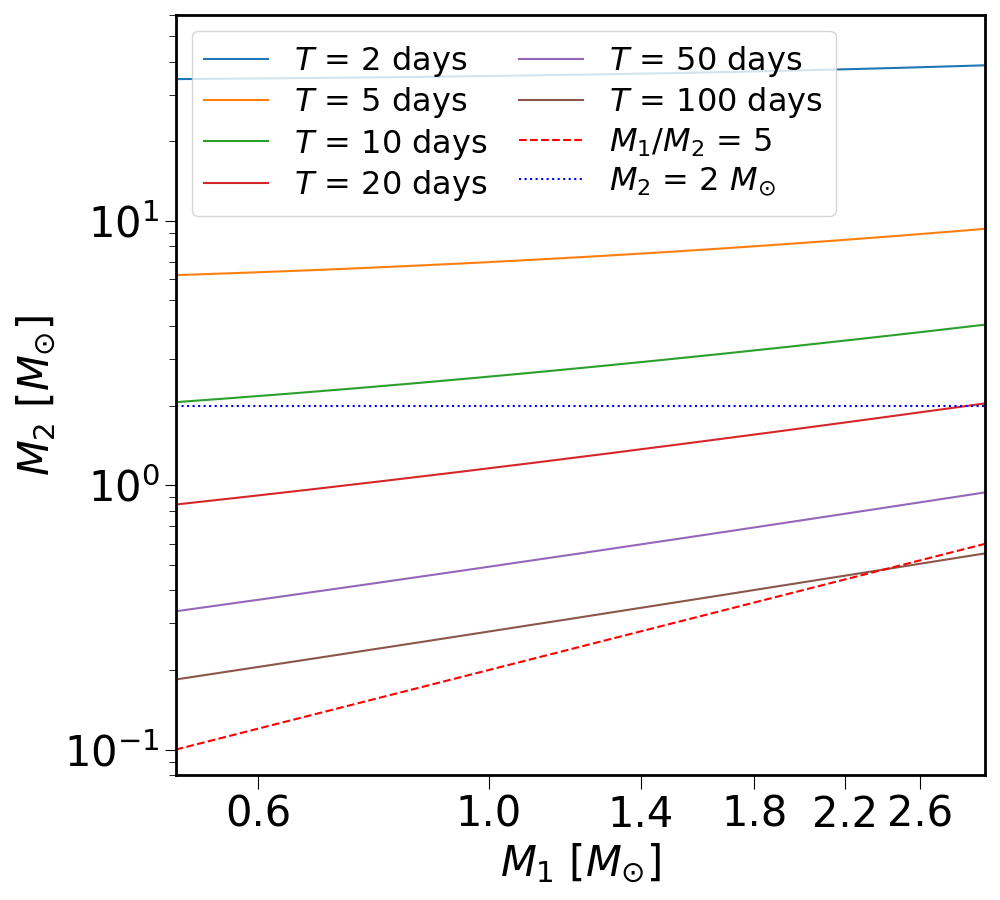}}    
    \subfloat[]{\includegraphics[height=0.3\textwidth]{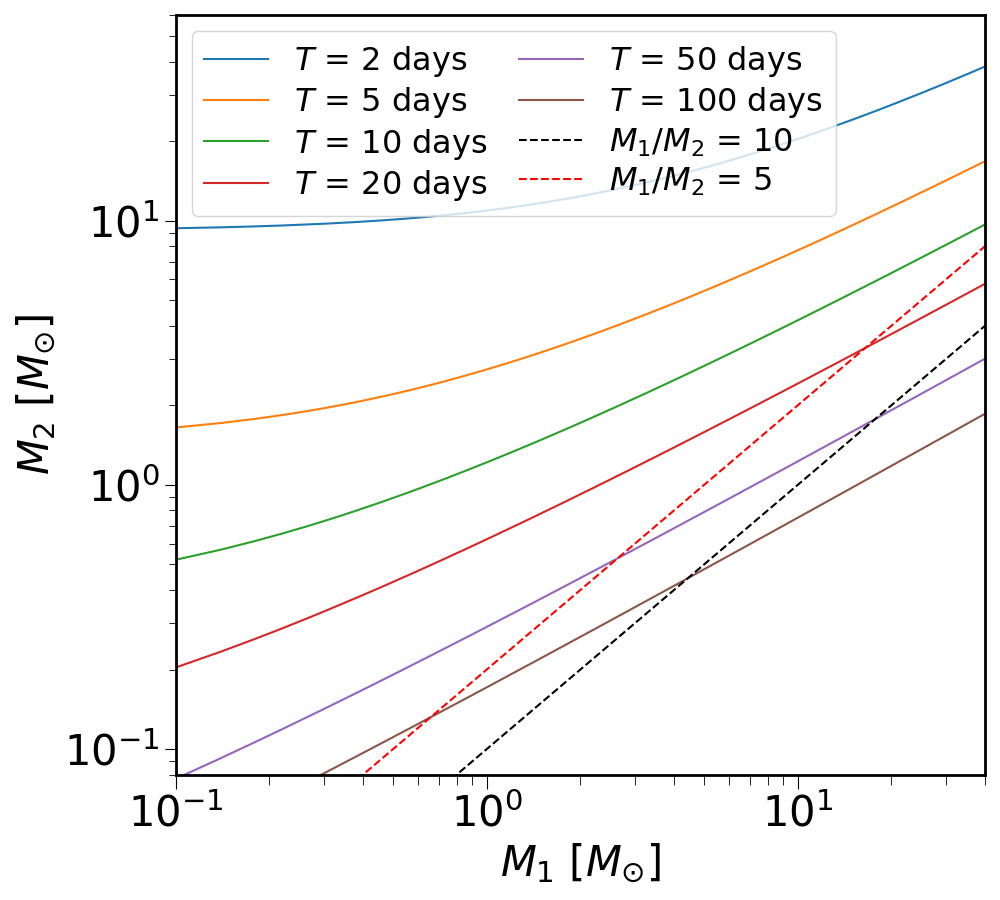}}
    \subfloat[]{\includegraphics[height=0.3\textwidth]{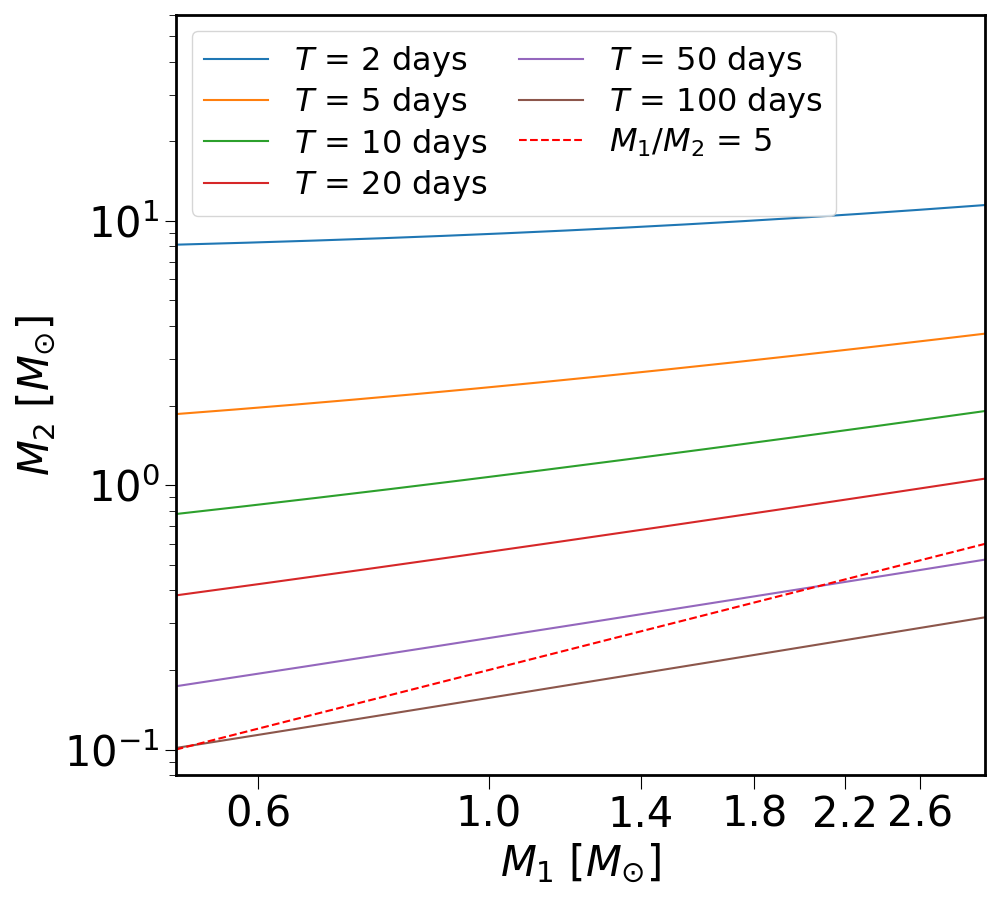}}
    \vspace{-0.4cm}
	\caption{Distribution of $M_1$ and $M_2$ at different $T$ for monitorable CBs (regions above respective solid lines). (a) BH-NSs, PSR binaries and rotation-powered tMSPs at $d$ = 5 kpc with PSR flux densities of $S = 0.5$ mJy (1.6 GHz). (b) BHXBs, NSXBs, AMXPs and novae at $d$ = 5 kpc and $S = 170$ $\mu$Jy (15 GHz). (c) accretion-powered tMSPs at $d$ = 2 kpc and $S = 75$ $\mu$Jy (15 GHz).}
    \label{fig:monitoring_binaries-compact-object-astrometry}
\end{figure*}

Figure \ref{fig:monitoring_binaries-compact-object-astrometry} presents the mass distributions of monitorable CBs with fixed $a_1$ using SKA-VLBI
\begin{equation}\label{equ:M1M2-distribution-compact-object-astrometry}
    M_1 \leq \sqrt{C} M_2^{3/2} - M_2, \; C =\frac{T^2 G}{4 \pi^2}\frac{1}{a_1^3}, \; a_1 = d \frac{\theta \Delta S}{2S},
\end{equation} 
where $d$ is less than that set by $\mathrm{DR_{max}}$ or the parallax limitation, and $G$ is the gravitational constant. SKA-VLBI will enable the development of detection and identification methodologies for BH-NSs as well as measurements of their orbital elements and component masses. It can monitor nearly all BH-NSs within 5 kpc with $T \geq 20$ days, as well as a subset of these systems with $T$ as short as several days. Similarly, this capability also applies to other CBs (e.g., BHXBs, NSXBs, AMXPs, and novae within 5 kpc, and accretion-powered tMSPs within 2 kpc), with more details provided in Figure \ref{fig:monitoring_binaries-compact-object-astrometry}.

Compact objects in high-mass XBs \citep{Fortin+2023, Ge+2024} MXB 0656-072 ($a_1 \sim 4.6$ mas, distance $\sim$0.2 kpc, and J2000 declinations (Decl.) $\sim${}$-7.2^{\circ}$) and 4U 0352+309 ($a_1 \sim 3.1$ mas, distance $\sim$0.6 kpc, and Decl. $\sim$31.0$^\circ$) could potentially be monitored by AA4, with MXB 0656-072 being potentially monitorable as early as AA2.

In summary, SKA-VLBI will acquire a sample of CBs including BH-NSs, monitor their orbital motion, and derive the key physical parameters such as orbital elements and component masses. These measurements will advance studies of: stellar structure and evolution under extreme conditions, the equation of state of nuclear matter, relativistic effects in strong gravitational fields, and formation of gravitational wave sources.

\section{Intermediate-mass Black Holes in Star Clusters of the Milky Way}\label{sec:IMBH-compact-object-astrometry}

Compelling IMBH candidates spread throughout the universe, including dwarf galaxies \citep[e.g.,][]{Yang+2020, Yang+2023}, nuclear star clusters, globular clusters, and young massive clusters \citep[YMCs; with ages $\sim$1--100 Myr and masses $\gtrsim$10$^{4}$ $M_{\odot}$;][]{Portegies-Zwart+2010}. Nevertheless, the jury is still out on whether they truly exist. The abundance of radio stars in YMCs, as exemplified by the Quintuplet cluster \citep[located at $\sim$8 kpc with an age of 3--5 Myr;][]{Gallego2022}, suggests that these clusters are favourable environments for investigating IMBH-dominated dynamics at radio wavelengths. Hundreds of young clusters have been identified in the Milky Way, among which several YMCs within several kpc (e.g., Westerlund 1, RSGC 1, Stephenson 2, the Arches cluster, and the Quintuplet cluster) are excellent candidates for hunting IMBHs.

\subsection{Radio Emissions Associated with IMBH Accretion}\label{sec:IMBH-YMC-emission-compact-object-astrometry}

Radio observations are important for determining whether the accretor is an IMBH, sBH, or NS \citep[e.g.,][]{Mezcua2015}, and thus for investigating the existence of IMBHs. In the absence of radio observations, we apply the Bondi accretion model combined with the empirical fundamental plane relation \citep[see equations (1)--(2) in][]{Karimi2024} to estimate the radio flux density at 5 GHz, $S_{\mathrm{5\; GHz}}$, associated with IMBH accretion. This relation is
\begin{equation}
    S_{\mathrm{5\; GHz}} \sim 2500 \left(\frac{M_{\mathrm{BH}}}{1000\; M_{\odot}}\right)^{1.98}\left(\frac{d}{1 \; \mathrm{kpc}}\right)^{-2}  \mathrm{mJy},
\end{equation}
where $M_{\mathrm{BH}}$ and $d$ denote IMBH mass and distance, respectively. 
For a 100 $M_{\odot}$ IMBH at 10 kpc, the predicted flux density (flat spectrum) will be $\sim$0.26 mJy. Even with AA2 observations, a 1-min integration achieves a sensitivity of $\sim$22 $\mu$Jy at 8.0 GHz, enabling $\sim$12$\sigma$ detection of radio emission from IMBHs.
Furthermore, accretion theories for IMBHs will be advanced, and crucial clues for accretion in other BH types will emerge, from mass measurements enabled by the SKA/SKA-VLBI.

Following the systematic search for IMBH candidates in star clusters, the investigation will be extended to include spectral index, variability, proper motions, polarization, jet morphology, and radio lobe structures. 

Similar investigations could target IMBHs in dwarf galaxies, which have also been widely considered as promising hosts of IMBHs \citep[e.g.][]{Yang+2020, Yang+2023}. Furthermore, SKA/SKA-VLBI will enhance the study of weak nuclear activity of dwarf galaxies, through the detection of faint, compact cores and the resolution of their nuclear structures on sub-parsec scales. 

\subsection{Monitoring Radio Stars Influenced by Possible IMBHs in Young Massive Star Clusters} \label{sec:monitoring-IMBH-compact-object-astrometry}

The IMBH candidate can be confirmed by astrometric monitoring of its bound radio stars to determine its dynamical mass. We employ a YMC (the Quintuplet cluster) paradigm to examine dynamic monitoring of radio stars surrounding IMBHs. The radio continuum emission from this YMC is dominated by massive (binary) stars \citep{Gallego2022}. Their median, maximum, and 5$\sigma$ flux densities are $\sim$8, $\sim$65, and $\sim$1 mJy/$d^2$ at 15 GHz, respectively, with average $\alpha \sim 0.6$. These stars are detected down to $\sim$1000 au from the cluster center, where the synthesized beam size corresponds to a physical scale of $\sim$2000 au. The sensitivity and spatial resolution limitations may preclude the detection of some radio stars. Radio stars are more detectable in nearer YMCs or with more sensitive and higher-resolution SKA/SKA-VLBI.

\begin{figure*}[ht!]
    \centering
    \vspace{-0.7cm}
    \subfloat[]{\hspace{-0.8 cm}\includegraphics[height=0.33\textwidth]{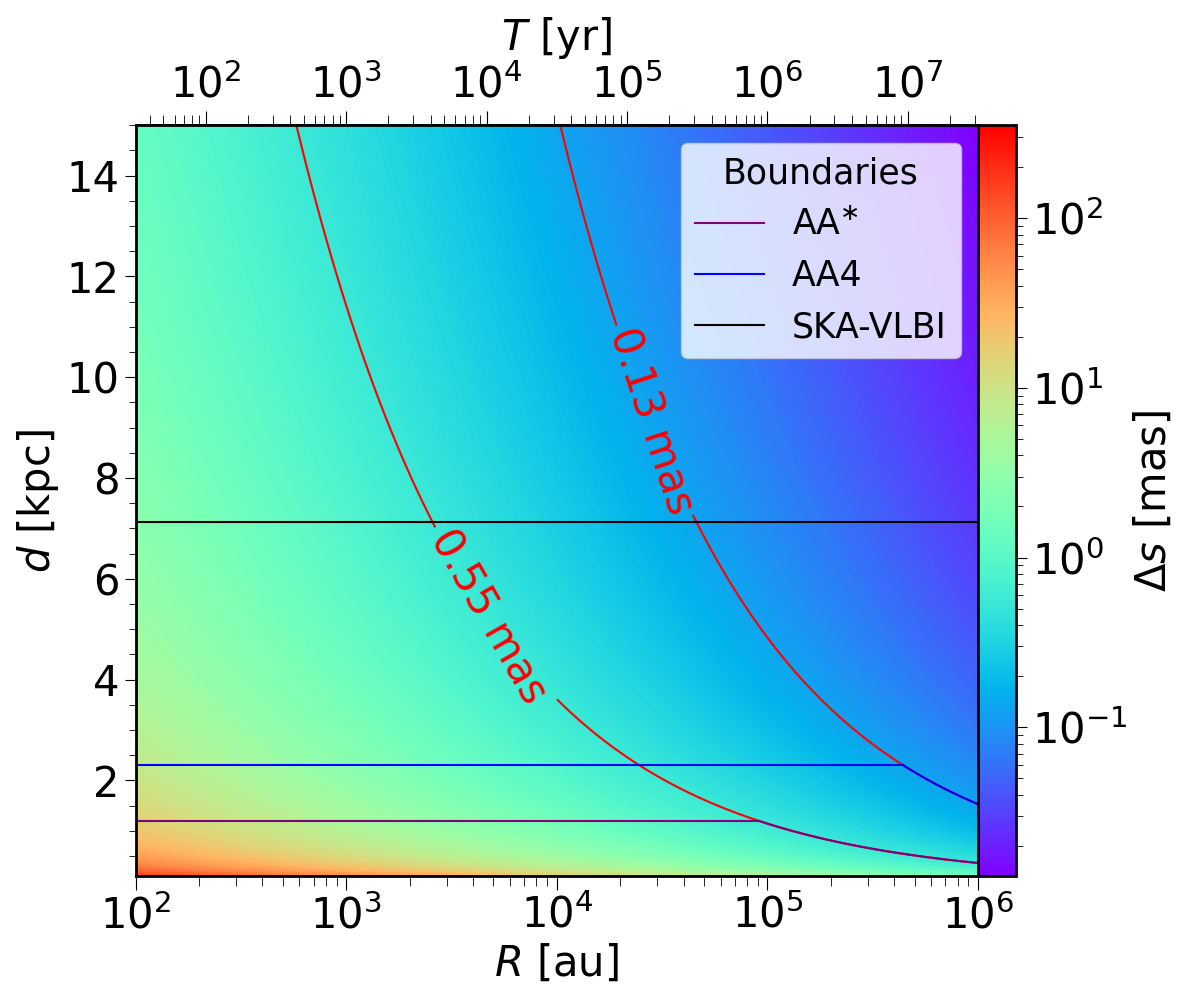}\hspace{1.6 cm}}
    \subfloat[]{\includegraphics[height=0.33\textwidth]{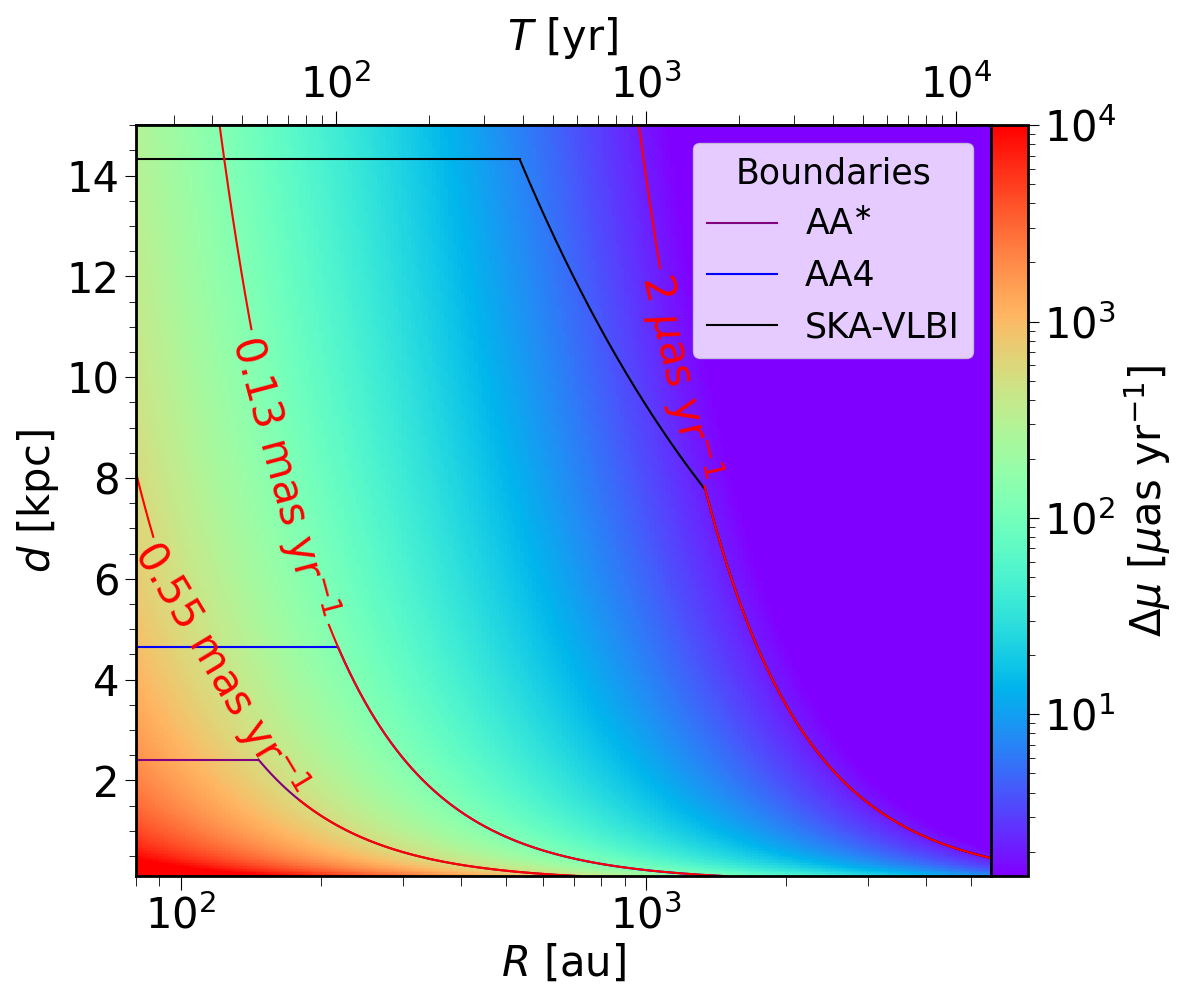}}
    \vspace{-0.4cm}
\caption{The distribution of $\Delta s$ (a) and $\Delta \mu$ (b) at different $R$ (or orbit period $T$) and $d$. Radio stars can be monitored by AA$^{\ast}$, AA4 or SKA-VLBI when they are below the purple, blue or black lines, respectively. The model adopts a 15 GHz observing frequency, a 10-min integration time, and flux densities, $S$, of (a) 8 and (b) 65 mJy/$d^2$. $\mathrm{DR_{max}}$ and a proper motion limit (with precision being $\theta/(2\cdot\mathrm{DR}\cdot \Delta t)$ and $\mathrm{DR} = S/\Delta S$) are incorporated in the calculations. \label{fig:motion-monitor-IMBH-compact-object-astrometry}}
\end{figure*}

For a radio star with orbital radius, $R$, the angular position change, $\Delta s$, and radial proper motion variation, $\Delta \mu$, over one year, $\Delta t = 1$ yr, are
\begin{equation}\label{equ:position-various-IMBH-compact-object-astrometry}
    \Delta s = \sqrt{\frac{GM_{\mathrm{BH}}}{R}} \frac{\Delta t}{d}, \; \; \; \;\;\; \;\;\;
    \Delta \mu =  \frac{GM_{\mathrm{BH}}}{R^2} \frac{\Delta t}{\kappa d},
\end{equation}
where $\kappa$ is the conversion factor from angular to linear velocity at a given distance. Figure \ref{fig:motion-monitor-IMBH-compact-object-astrometry} presents the $\Delta s$ and $\Delta \mu$ of a radio star orbiting an IMBH of 1000 $M_{\odot}$. Even AA$^{\ast}$ and AA4 will be capable of monitoring $\Delta s$ of radio stars influenced by IMBHs in the solar neighborhood at 15 GHz with 10-min integration, while SKA-VLBI could extend this capability to systems $\sim$7 kpc away. However, measuring $\Delta \mu$ remains challenging. SKA-VLBI may achieve $\Delta \mu$ measurements for bright radio stars (e.g., $\sim$65 mJy/$d^2$) in IMBH systems, thereby enhancing dynamical studies. The primary limiting factor is sensitivity, e.g., the detectable distance extends to $\sim$10 kpc for $\Delta s$ monitoring at 1-hr integration. These results demonstrate that SKA/SKA-VLBI will enable dynamical mass measurements of IMBHs through multi-stellar tracking.

\subsection{Possible Central IMBHs in Other Clusters}\label{sec:other-cluster-IMBH-compact-object-astrometry}

Beyond YMCs, it may also be feasible to investigate central IMBHs in other types of star clusters through radio observations. For IMBHs in globular clusters with dozens of PSRs, the approach is analogous to that for IMBH systems in YMCs, and the detectable distance for these IMBHs is reduced by $\sim$30-40\%. In globular or open clusters with sparse PSR populations, it may be necessary to monitor one or a few PSRs that have traversed more than half of their orbital period. For a central IMBH of 1000 $M_{\odot}$, a 10 yr orbital period corresponds to an orbital radius of $\sim$50 au. With a 10-min integration, SKA-VLBI could resolve the orbital motion of a PSR at this 50 au separation in a system within $\lesssim$7 kpc.

In summary, SKA/SKA-VLBI are expected to achieve the first conclusive identifications of IMBHs and measure their dynamical masses, providing critical constraints on BH formation and evolution, and advancing gravitational wave astronomy.

\section{Summary and Conclusion}\label{sec:summary-compact-object-astrometry}

This work outlines research prospects for studying compact objects using the high sensitivity ($\mu$Jy level) and ultrahigh-precise astrometry ($\mu$as level) of SKA/SKA-VLBI, with emphasis on potential breakthroughs in compact binaries (CBs) and intermediate-mass black holes (IMBHs). The SKA-Mid AA2-AA4 surveys are expected to systematically identify numerous CB and IMBH candidates. Orbital motion monitoring with SKA/SKA-VLBI will transform our understanding of compact object physics research by: (1) Precise determination of CB orbital elements and masses, testing stellar structure and evolution under extreme conditions; (2) Developing effective methodology for black hole-neutron star binary (BH-NS) identification to catalog binaries and probe nuclear matter equation of state, strong-field relativistic effects, and gravitational-wave source formation; (3) Definitive IMBH identification and mass measurements to constrain BH formation and evolution, advancing gravitational-wave astronomy.

\textbf{\large Acknowledgements} 

We thank the anonymous referee for the helpful comments and suggestions. This work is supported by the National SKA Program of China (grant No. 2022SKA0120103), the NSFC Grants Nos. 12203104, 12403077 and 12403041, and the Key Laboratory for Radio Astronomy. 

\bibliographystyle{abbrvnat-maxbibnames4}
\bibliography{Compact_objects} 

\end{CJK*}
\end{document}

%% file: journal-names.tex
\newcommand{\actaa}{Acta Astron.} 
\newcommand{\araa}{Annu. Rev. Astron. Astrophys.} 
\newcommand{\aar}{Astron. Astrophys. Rev.} 
\newcommand{\ab}{Astrobiol.} 
\newcommand{\aj}{Astron. J.} 
\newcommand{\apj}{Astrophys. J.} 
\newcommand{\apjl}{Astrophys. J. Lett.} 
\newcommand{\apjs}{Astrophys. J. Suppl. Ser.} 
\newcommand{\ao}{Appl. Opt.} 
\newcommand{\apss}{Astrophys. Space Sci.} 
\newcommand{\aap}{Astron. Astrophys.} 
\newcommand{\aapr}{Astron. Astrophys. Rev.} 
\newcommand{\aaps}{Astron. Astrophys. Suppl.} 
\newcommand{\baas}{Bull. Am. Astron. Soc.} 
\newcommand{\caa}{Chinese Astron. Astrophys.} 
\newcommand{\cjaa}{Chinese J. Astron. Astrophys.} 
\newcommand{\cqg}{Class. Quantum Gravity} 
\newcommand{\gal}{Galaxies} 
\newcommand{\gca}{Geochim. Cosmochim. Acta} 
\newcommand{\icarus}{Icarus} 
\newcommand{\jcap}{J. Cosmol. Astropart. Phys.} 
\newcommand{\jgr}{J. Geophys. Res.} 
\newcommand{\jgrp}{J. Geophys. Res.: Planets} 
\newcommand{\jqsrt}{J. Quant. Spectrosc. Radiat. Transf.} 
\newcommand{\memsai}{Mem. Soc. Astron. Italiana} 
\newcommand{\mnras}{Mon. Not. R. Astron. Soc.} 
\newcommand{\nat}{Nature} 
\newcommand{\nastro}{Nat. Astron.} 
\newcommand{\ncomms}{Nat. Commun.} 
\newcommand{\nphys}{Nat. Phys.} 
\newcommand{\na}{New Astron.} 
\newcommand{\nar}{New Astron. Rev.} 
\newcommand{\physrep}{Phys. Rep.} 
\newcommand{\pra}{Phys. Rev. A} 
\newcommand{\prb}{Phys. Rev. B} 
\newcommand{\prc}{Phys. Rev. C} 
\newcommand{\prd}{Phys. Rev. D} 
\newcommand{\pre}{Phys. Rev. E} 
\newcommand{\prl}{Phys. Rev. Lett.} 
\newcommand{\psj}{Planet. Sci. J.} 
\newcommand{\planss}{Planet. Space Sci.} 
\newcommand{\pnas}{Proc. Natl Acad. Sci. USA} 
\newcommand{\procspie}{Proc. SPIE} 
\newcommand{\pasa}{Publ. Astron. Soc. Aust.} 
\newcommand{\pasj}{Publ. Astron. Soc. Jpn} 
\newcommand{\pasp}{Publ. Astron. Soc. Pac.} 
\newcommand{\rmxaa}{Rev. Mexicana Astron. Astrofis.} 
\newcommand{\sci}{Science} 
\newcommand{\sciadv}{Sci. Adv.} 
\newcommand{\solphys}{Sol. Phys.} 
\newcommand{\sovast}{Soviet Ast.} 
\newcommand{\ssr}{Space Sci. Rev.} 
\newcommand{\uni}{Universe} 
\newcommand{\prx}{Phys. Rev. X} 
\newcommand{\lrr}{Living Rev. Relativ.} %
\newcommand{\raa}{Res. Astron. Astrophys.} %